\documentclass[aps,prl,amsfonts,twocolumn,superscriptaddress,showpacs,floatfix]{revtex4}

\usepackage{graphicx}
\usepackage{amsmath}
\usepackage{bm}

\begin{document}

\def\be{\begin{equation}}
\def\ee{\end{equation}}
\newcommand{\corr}[1]{\langle #1\rangle}
\newcommand{\ccorr}[1]{\langle\langle #1\rangle\rangle}
\def\var{\mathop{\rm var}}
\def\std{\mathop{\rm std}}
\def\bp{{\bf p}}
\def\bq{{\bf q}}
\def\br{{\bf r}}
\def\brho{\bm{\rho}}
\def\eps{\varepsilon}
\def\tr{\mathop{\rm tr}}

\title{Quantum percolation in granular metals}

\author{M. V. Feigel'man}
\affiliation{L. D. Landau Institute for Theoretical
Physics, Moscow 119334, Russia}
\affiliation{Materials Science Division, Argonne National Laboratory,
Argonne, Illinois 60439, USA}

\author{A. S. Ioselevich}
\affiliation{L. D. Landau Institute for Theoretical Physics,
Moscow 119334, Russia}

\author{M. A. Skvortsov}
\email{skvor@itp.ac.ru}
\affiliation{L. D. Landau Institute for Theoretical Physics,
Moscow 119334, Russia}
\affiliation{Materials Science Division, Argonne National Laboratory,
Argonne, Illinois 60439, USA}

\date{April 14, 2004}
\begin{abstract}
Theory of quantum corrections to conductivity of granular metal films
is developed for the realistic case of large randomly distributed
tunnel conductances. Quantum fluctuations of intergrain voltages
(at energies $E$ much below bare charging energy scale $E_C$)
suppress the mean conductance $\overline{g}(E)$ much stronger than its
standard deviation $\sigma(E)$. At sufficiently low energies $E_*$ any
distribution becomes broad, with $\sigma(E_*) \sim \overline{g}(E_*)$,
leading to strong local fluctuations of the tunneling density of states.
Percolative nature of metal-insulator transition is established by
combination of analytic and numerical analysis of the matrix
renormalization group equations.
\end{abstract}

\pacs{71.30.+h, 64.60.Ak, 73.23.Hk}

%71.30.+h   Metal-insulator transitions and other electronic transitions
%64.60.Ak   Renormalization-group, fractal, and percolation studies
%           of phase transitions
%73.23.Hk   Coulomb blockade; single-electron tunneling

\maketitle

{\em Introduction.}---%
Low-temperature electron transport in granular metals was
intensively studied during last
years~\cite{efetov,bel1,bel2,kam04}. It was shown that in the
temperature range $T \geq g\delta$ (where $g \gg 1$ is
characteristic value of dimensionless intergrain conductance in
units of $e^2/2\pi\hbar$ and $\delta$ is the intragrain level
spacing), quantum corrections to conductivity originate mainly
from local fluctuations of voltages between neighboring grains.
This effect can be treated within Ambegaokar-Eckern-Schoen
model~\cite{AES} and leads to logarithmic temperature dependence
of the effective conductance~\cite{efetov}:
\be
  g(T) = g_0 - \frac{4}{z}\ln \frac{g_0E_C}{T},
\label{1}
\ee
where $E_C \gg \delta$ is the charging energy of an
individual grain, $g_0$ is the bare tunneling conductance of
intergrain junctions (identical for all junctions), and $z$ is the
coordination number of the lattice~\cite{comment-g}. The result
(\ref{1}) is valid as long as the renormalized conductance $g(T)$ is
large, i.e. down to temperatures $T_1 = g_0E_Ce^{-zg_0/4}$.
It was argued~\cite{bel1,bel2} that transition from metal
to insulator behavior (MIT, for brevity) occurs at $T\sim T_1$ as long as
$T_1 \geq g\delta$. The same conclusion for the two-dimensional (2D)
case was reached~\cite{kam04} via instanton analysis.

Although the above  results may well be applied to artificial 2D
arrays of well-defined tunnel junctions, tunnel conductances
$g_{ij}$ are random in granular metals.
In this Letter we investigate the role of $g_{ij}$ randomness
for energy (temperature) dependent properties of thin granular films,
such as macroscopic conductance $g_{\text{eff}}(T)$ and the local
tunneling density of states (LTDoS) $\nu_i(E)$.
Quantum fluctuations lead to suppression of $g_{ij}$
described by the one-loop renormalization group (RG) equation:
\be
\label{RGE}
  \frac{d g_{ij}}{dt} = - 2g_{ij} R_{ij},
\ee
where $t(E)=\ln(\overline{g}_0E_C/E)$ is the auxiliary RG ``time",
$\overline{g}_0$ being some mean bare conductance,
and $R_{ij}$ is the resistance of the network between the
points $i$ and $j$. Physically, renormalization of $g_{ij}$ is due
to fluctuations of voltage between the grains $i$ and $j$, which are
governed by the corresponding resistance $R_{ij}$. Eq.~(\ref{RGE})
is a straightforward generalization of the RG equation for a
regular array~\cite{efetov} with $g_{ij} \equiv g$ and $R_{ij} =
2/g z$, whose solution is given by Eq.~(\ref{1}). The system of RG
equations (\ref{RGE}) is nontrivial since $R_{ij}$ is a
complicated nonlocal function of all individual conductances
$g_{kl}$.

\begin{figure}
\includegraphics[width=0.9\columnwidth]{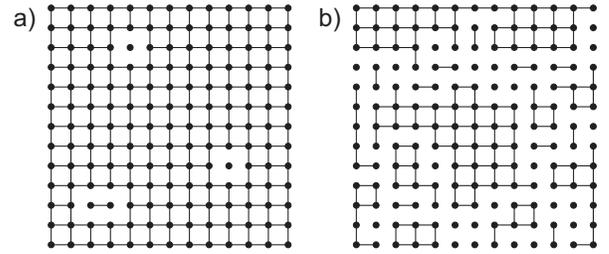}
\caption{Conducting bonds (with $g_{ij}>0$) at different values of $t$.
Results of numerical simulation of Eq.~(\protect\ref{RGE})
on the lattice $20\times20$ with
$P_0(g)=(2g_0\theta(g)/\pi)/(g^2+g_0^2)$.
a) $t=0.64g_0$ with the fraction of conducting bonds $N_{\text{cond}}=0.95$;
b) $t=1.08g_0$ with $N_{\text{cond}}=0.55$.}
\label{F:clusters}
\end{figure}

In a regular system, Eq.~(\ref{RGE}) drives all conductances to zero
simultaneously at $t=t_c=zg_0/4$, marking the point of the MIT.
We will show that in a random system
renormalized conductances of the junctions collapse to zero
neither all at once, nor one by one, but in groups.
These groups enclose clusters, consisting of one
or several sites, which become disconnected from the rest
of the network after the collapse (see Fig.~\ref{F:clusters}).
As a result, the MIT in a natural granular system is a percolative
transition: it takes place, when enough clusters have become
disconnected so that the percolation via still conducting links
is destroyed.

The above picture of conductances, eventually collapsing to zero,
follows from the one-loop RG equation (\ref{RGE}).
The one-loop approximation breaks down at $g\sim 1$; at lower energies
the conductance decays exponentially with the RG ``time" $t(E)$.
Therefore, Eq.~(\ref{RGE}) can adequately describe
only the initial stage of the MIT.
Nevertheless, there exists a region near the
transition where the percolative cluster contains
good conductances with $g>1$ so that Eq.~(\ref{RGE}) is still applicable.

%One should have in mind, however,
%that the one-loop approximation breaks down at $g\lesssim 1$. Thus,
%when we speak about the collapse, we actually mean that corresponding conductances
%become of order of unity. Present results adequately describe only the initial stage of MIT,
%where the effective global conductivity
%drops from $g_{\rm eff}\sim \overline{g}_0\gg 1$ down to $g_{\rm eff}\sim 1$, while the final stage,
%where $g_{\rm eff}$ ultimately  drops  to zero, remains a puzzle.

We start from the case of
relatively narrow original distribution $P_0(g)$ characterized by
the mean value $\overline{g}_0$ and the standard deviation
$\sigma_0 \ll \overline{g}_0$, and show that the renormalized
distribution $P(g)$ broadens. In particular, for the square
lattice ($z=4$):
\be
  \frac{\sigma(E)}{\overline{g}(E)}
 = \frac{\sigma_0}{\overline{g}_0}\frac{\overline{g}_0/\overline{g}}
 {\sqrt{2\ln(\overline{g}_0/\overline{g})\ln\ln(\overline{g}_0/\overline{g})}},
\label{2}
\ee
%\be
%  \frac{\sigma(E)}{\overline{g}(E)}
% = \frac{\sigma_0}{\overline{g}_0}\frac{e^s}{\sqrt{2s\ln s}} ,
%\label{2} \ee
where $\overline{g}\equiv \overline{g}(E)
=\overline{g}_0-\ln(\overline{g}_0E_C/E)$.
%where $s(E) = -\ln \left( 1-\frac1{\overline{g}_0}\ln\frac{\overline{g} _0E_C}{E} \right)$.
Eq.~(\ref{2}) is a large-$\ln(\overline{g}_0/\overline{g})$ asymptotics of a more general
expression (see Eq.~(\ref{s/s}) below).
It is valid as long as $\sigma(E)\ll\overline{g}(E)$, i.e.\
above
%up to $s = s_* = \ln(\overline{g}_0/\sigma_0) +
%\frac12\ln\ln(\overline{g}_0/\sigma_0)$.
% where $\sigma(E_*)/\overline{g}(E_*) \approx 1$.
%The corresponding energy/temperature scale
$E_*=T_* = \overline{g}_0E_Ce^{-\overline{g}_0 + \sigma_0} \gg T_1$,
where $T_1$ marks MIT in an ideal array with $\sigma_0 \ll 1$. Thus
transition from metal into insulator in a granular array is
intrinsically inhomogeneous. The vicinity of this transition at $
\max(E,T) \leq T_*$ is difficult for exact analytical treatment
as the width of distribution $P(g|E)$ becomes of order of its mean value. In
this region we employ the effective-medium approximation (EMA)
and numerical solution of the RG equations (\ref{RGE}),
and demonstrate that MIT is of percolative nature.

Strong self-developed inhomogeneity of a granular array can be
probed by scanning tunneling measurement of the LTDoS modified
by the Coulomb zero-bias anomaly (ZBA) \cite{AA,Fin,LS}.
ZBA modification $Z(E) = \nu(E)/\nu_0$ of the average LTDoS in a regular
array was considered in Refs.~\cite{efetov,bel2} and found to
become very large before approaching MIT. Here we analyze spatial
fluctuations of the ZBA suppression factor $Z_i(E)$. For an
originally narrow distribution $P_0(g)$, the log-normal
distribution of the ZBA factors is found, with $\std[\ln Z_i(E)]
\approx \sigma(E)/\overline{g}(E)$. Thus we predict order-of-unity
local fluctuations of LTDoS at $\max(E,T) \leq T_*$. Spatial
correlation length $\xi(E)$ of these fluctuations was found to
grow moderately with $E$ decrease in the case of weak original
disorder: $\xi(E) \approx  \sqrt{\ln[\overline{g}_0/\overline{g}(E)]}$,
reaching $\sqrt{\ln(\overline{g}_0/\sigma_0)}$ at the border of strong
inhomogeneity $E \sim T_*$. For the region  in the
vicinity of MIT, where relative fluctuations are large,
we present numerical analysis of LTDoS fluctuations.
Below we provide brief derivation of our results.

{\em Narrow distribution.}---%
If the standard deviation $\sigma$ of the distribution is much
smaller than the mean $\overline{g}$, the latter follows the
homogeneous solution (\ref{1}): $\overline{g}(t)=\overline{g}_0-t$, while
evolution of $\delta g_{ij}=g_{ij}-\overline{g}$ can be described
perturbatively.
Resistance can be written as $R_{ij}=G_{ii} + G_{jj} - 2G_{ij}$,
where $G_{ij} = \hat{A}^{-1}_{ij}$ is the Green function
of the diffusion operator on the network
defined by the matrix elements $A_{ii}=\sum_j g_{ij}$ and
$A_{ij}=-g_{ij}$ \cite{comment-A}.
Using the standard perturbative series $G_{ij} =
\overline{G}_{ij} - \overline{G}_{ik} \delta A_{kl} \overline{G}_{lj} + \dots$
we find
\be
  \delta G_{ij}
= - \sum_{\corr{kl}} \delta g_{kl} (\overline{G}_{ik}-\overline{G}_{il})
(\overline{G}_{jk}-\overline{G}_{jl}) ,
\label{dG}
\ee
where in momentum
representation $\overline{G}(\bp)=[2\overline{g}\eps(\bp)]^{-1}$;
for the square lattice, $\eps(\bp) = 2 - \cos p_x - \cos p_y$.

To proceed further we choose a vector representation for
conductances $g_i^\alpha$ when each edge is characterized by the
lattice site $i$ it goes from and direction $\alpha$ which can be
either horizontal ($+x$) or vertical ($+y$).
Using Eq.~(\ref{dG}), introducing a new
time variable $s = \ln[\overline{g}_0/\overline{g}(t)] =
-\ln[1-t/\overline{g}_0]$
and passing to Fourier representation we get
a linear evolution:
\be
  \frac{d\delta g_\alpha(\bp)}{ds}
  = - \mathfrak{M}_{\alpha\beta}(\bp) \delta g_\beta(\bp) ,
\ee
governed by the $2\times2$ time-independent matrix
\be
  \mathfrak{M}(\bp)
  = 1 - {\cal P}_1(\bp)
  - {\cal P}_2(\bp)
  (e^{i\alpha_\bp}\hat\sigma_+ + e^{-i\alpha_\bp}\hat\sigma_-) ,
\label{M}
\ee
where $\alpha_\bp=(p_x-p_y)/2$ and
$\hat\sigma_\pm=(\hat\sigma_1\mp\hat\sigma_2)/2$, $\hat\sigma_k$
being the Pauli matrices. The functions ${\cal P}_{1,2}(\bp)$ are given
by
\begin{subequations}
\label{P(p)}
\begin{gather}
  {\cal P}_1
  = 2 \int (d\bq)
   \frac{(1-\cos q_x)(1-\cos(p_x-q_x))}
  {\eps(\bq)\eps(\bp-\bq)} ,
\label{P1}
\\
  {\cal P}_2
  = 2
  \int (d\bq)
  \frac{(\cos\frac{p_x}{2}-\cos q_x)(\cos\frac{p_y}{2}-\cos q_y)}
  {\eps(\bq+\bp/2)\eps(\bq-\bp/2)} ,
\label{P2}
\end{gather}
\end{subequations}
where the integral with $(d\bq) \equiv d^2q/(2\pi)^2$ runs over
the Brillouin zone. At small $\bp$ they have a nonanalytic behavior:
${\cal P}_1(\bp)=1-1/\pi-(p^2/8\pi)\ln(1/p)+\dots$ and
${\cal P}_2(\bp)=1/\pi-(p^2/8\pi)\ln(1/p)+\dots$

The eigenvalues of the matrix (\ref{M}) form two branches:
\be
  \lambda_\pm(\bp) = 1 - {\cal P}_1(\bp) \pm {\cal P}_2(\bp),
\ee
the eigenfunctions being $(e^{i\alpha_\bp}, \mp1)^T/\sqrt{2}$.
The spectral branch $\lambda_+(\bp)$ is gapped whereas the branch
$\lambda_-(\bp)$ becomes gapless in the long wave-length limit:
$\lambda_-(\bp\to0)\approx (p^2/4\pi)\ln(1/p)$. Once the spectral
properties of the matrix $\mathfrak{M}$ are known one can express
$\delta g$'s at time $s$ via their initial values at $s=0$:
\be
 \delta g_\alpha(\br,s)
 = \sum_{\br'} K_{\alpha\beta}(\br-\br',s) \delta g_\beta(\br',0) .
\label{evol-r} \ee The Fourier-transformed kernel is given by
\be
 K(\bp,s)
 = K_1(\bp,s)
 + K_2(\bp,s) (e^{i\alpha_\bp}\hat\sigma_++e^{-i\alpha_\bp}\hat\sigma_-) ,
\label{K(p)} \ee
where $K_{1,2}(\bp,s)=(e^{-\lambda_-(\bp)s} \pm
e^{-\lambda_+(\bp)s})/2$.

Equation~(\ref{evol-r}) allows to find the evolution of the
single-site distribution function $P(g)$. A more convenient
quantity is the characteristic function $\chi(\lambda)$ defined as
$e^{-\chi(\lambda)} = \int P(\delta g) e^{i\lambda \delta
g} d\delta g$. Assuming that at $s=0$ different conductances are
uncorrelated we find for $\chi(\lambda)$:
$\chi(\lambda;s) =
  \sum_{\br}
  \bigl( \chi[K_1(\br,s)\lambda;0] + \chi[K_2(\br,s)\lambda;0] \bigr)$.
Thus, the variance of the distribution will decay as
\be
  \frac{\sigma^2(s)}{\sigma^2_0}
  =
  \frac12 \int (d\bp)
  \left[
    e^{-2\lambda_-(\bp)s} + e^{-2\lambda_+(\bp)s}
  \right] .
\label{s/s}
\ee

\begin{figure}
\includegraphics[width=0.9\columnwidth]{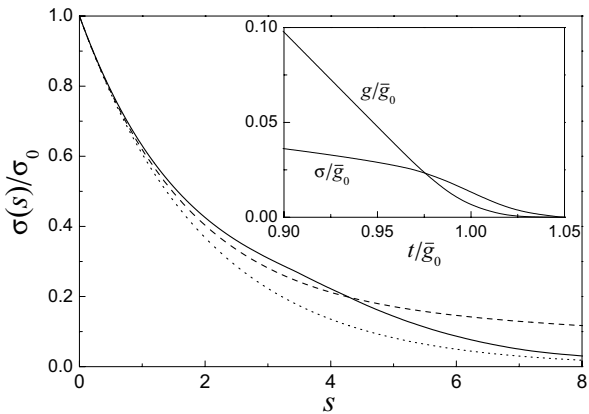}
\caption{Evolution of $\sigma(s)/\sigma_0$, Eq.~(\protect\ref{s/s}),
(dashed line), the result of numerical simulation of Eq.~(\protect\ref{RGE})
with $\sigma_0/\overline{g}_0=0.1$ (solid line),
and the EMA prediction $e^{-s/2}$ (dotted line).
Inset: $\overline{g}(t)$ and $\sigma(t)$ near the MIT.}
\label{F:s}
\end{figure}

For $s\ll1$, $\sigma(s)/\sigma_0 = 1 - s/2 + \dots$ This series is
to be compared with decay of the mean conductance:
$\overline{g}(s)/\overline{g}_0 = e^{-s} = 1 - s + \dots$. Thus, even at the
initial stage of the evolution the width of the distribution
decays slower than its average. In the case $s\gg1$ the integral
(\ref{s/s}) is dominated by the soft mode $\lambda_-(\bp)$ at
$\bp\to0$ leading to $\sigma^2(s)/\sigma^2_0 \approx 1/(2s\ln s)$
and hence to Eq.~(\ref{2}).
Practically, the applicability of this asymptotics
is limited to very large $s$.
For intermediate values of $s$ one has to employ full Eq.~(\ref{s/s}) with
numerical integration over the Brillouin zone.
The obtained function $\sigma(s)/\sigma_0$
together with the prediction of the EMA and results
of numerical simulation of Eq.~(\ref{RGE}) [for a typical
realization of disorder on the $20\times20$ lattice
with periodically boundary conditions
and $P_0(g)=(2g_0\theta(g)/\pi)/(g^2+g_0^2)$] is shown in Fig.~\ref{F:s}.

Apart from broadening the single-site distribution
$P(g)$, the RG flow (\ref{RGE}) produces correlations
between $\delta g$ at different links:
%\be
$
  C_{\alpha\beta}(\br;s_1,s_2)
  = \corr{\delta g_\alpha(\br,s_1) \, \delta g_\beta(0,s_2)}
$.
%\label{C(r)}
%\ee
The Fourier transform of the correlation function reads:
\be
  C(\bp;s_1,s_2) = \sigma^2_0 K(\bp;s_1+s_2) .
\label{C(p)}
\ee
At the initial stage of evolution, at $s\equiv
(s_1+s_2)/2\lesssim1$, correlations are short-ranged. At the
later stage, $s\gtrsim1$, correlations with large correlation
length $\xi(s) = \sqrt{(4/\pi)s\ln s}$ develop:
$C_{\alpha\beta}(\br,s_1,s_2)=\sigma^2(s)\exp[-r^2/\xi^2(s)]$.

Spatial fluctuations of $g_{ij}$ lead to fluctuations of the LTDoS
$\nu_i(E)=Z_i(E)\nu_0$. The ZBA suppression factor $Z_i(E)$  for
granular media at $E \geq g\delta $ can be found according to
simple ``environmental theory"~\cite{SET}:
\be
  \ln Z_i(E) = - 2\int_0^t R_i(t') dt' ,
\ee
where $R_i(t)$ is the resistance between the site $i$ and the
far region of the array at the energy scale $E=\overline{g}_0E_Ce^{-t}$.
The same result follows from the analysis provided in~\cite{efetov,bel2}.
It is important to note that in a homogeneously disordered metal
the short-length cutoff in the integral that determines the effective
resistance $R(E)$ is given by the diffusion length $\sqrt{\hbar D/E}$,
whereas in the present case it is just the grain size. The long-scale
cutoff for logarithmic divergency of $R_i(E)$ in 2D is $L(E) =
e^2 g_{\rm eff}/E$ (in the absence of external screening).
Thus one can write $R_i(E)=G_{ii}^{\text{reg}}$, where the
otherwise divergent $G_{ii}$ is regularized by the finite
length $L(E)$. Local fluctuations of $\ln Z_i(E)$ are determined
by a much smaller region of the size $\xi$ around the site $i$ so
that the object $\delta R_i=\delta G_{ii}$ is already free of
infra-red divergency and is independent on the details of
screening. Employing Eq.~(\ref{dG}) we obtain
\be
  \delta \ln Z_i(t)
  = - 2\int_0^t dt' \sum_{k,\alpha}
  \frac{\delta g_\alpha(\br_k,t')}{g_0^2(t')} Q_\alpha(\br_i-\br_k) ,
\ee
where $Q_\alpha(\br)$ is specified by its Fourier transform:
\be
 Q_\alpha(\bp)
 = e^{-ip_\alpha/2}
 \int (d\bq)
  \frac{\cos(p_\alpha/2)-\cos q_\alpha}
  {2\eps(\bq+\bp/2)\eps(\bq-\bp/2)} .
\label{Q(p)} \ee
Averaging $\delta \ln Z_i(t)$ with the help of
Eq.~(\ref{C(p)}), and changing integration variable
from $t$ to $s$ we get for the variance of the ZBA exponent:
\begin{multline}
\label{ZZ}
 \corr{[\delta \ln Z_i(t)]^2}
 = 4  \frac{\sigma_0^2}{\overline{g}_0^2}
  \int_0^s\!\!\int_0^s ds_1 ds_2 \, e^{s_1+s_2}
\\
{} \times
   \int (d\bp)
   K_{\alpha\beta}(\bp;s_1+s_2) Q_\alpha(\bp) Q_\beta(-\bp) .
\end{multline}

\begin{figure}
\includegraphics[width=0.9\columnwidth]{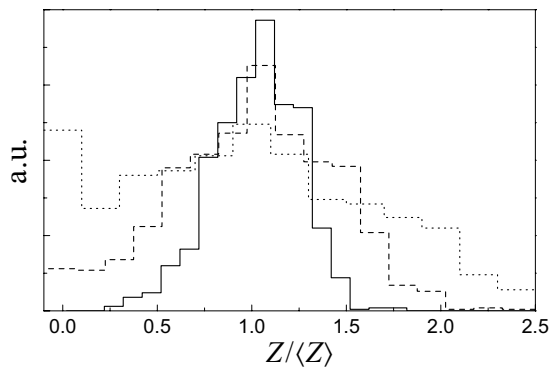}
\caption{Histograms for the distribution of the ZBA factors $Z(E)$
near the percolation threshold $t_c=0.99\,\overline{g}_0$ obtained
numerically for $\sigma_0/\overline{g}_0=0.32$:
$t=0.8\,t_c$ (solid line);
$t=0.9\,t_c$ (dashed line);
$t=0.95\,t_c$ (dotted line).}
\label{F:ZBA}
\end{figure}

At the initial stage of the evolution, for $s\lesssim1$,
$\corr{[\delta \ln Z_i(t)]^2} \approx
(0.26\, \sigma_0/\overline{g}_0)^2s^2$. In the region of well-developed
correlations, for $s\gtrsim1$, Eq.~(\ref{ZZ}) is dominated by
small momenta. In this limit Eq.~(\ref{Q(p)}) can be estimated
as $Q_\alpha(\bp\to0) = (1/4\pi) \ln(1/p)$,
yielding $\corr{[\delta \ln Z_i(t)]^2} \approx
(\sigma^2_0/8\pi^2\overline{g}^2_0) (e^{2s}\ln s/s)$. Rewriting this
result as $\corr{[\delta \ln Z_i(t)]^2} \approx (\ln s/2\pi)^2
\sigma^2(s)/\overline{g}^2(s)$ we see that fluctuations
of the ZBA factors $Z_i$ become of the order of unity simultaneously
with the renormalized ratio $\sigma/\overline{g}$.

The results of numerical simulation for a model distribution
$P(g)=\exp[-(\ln g/\overline{g}_0)^2/2\sigma_1^2]/(\sqrt{2\pi}\sigma_1g)$
with the moderately small variance
$\sigma_0^2=\overline{g}_0^2[e^{2\sigma_1^2}-e^{\sigma_1^2}]=(0.32\,\overline{g}_0)^2$
on the lattice $32\times32$ are shown in Fig.~\ref{F:ZBA},
where we present the distribution of the local values of $Z_i(E)$
at three values of the RG ``time'' $t$.
Upon lowering the energy scale and approaching
the MIT transition at $E \sim T_c=\overline{g}_0E_Ce^{-t_c}$
with $t_c=0.99\,\overline{g}_0$, we observe
a growing relative width of $Z$ distribution,
with the zero-$Z$ peak developing near the percolation
threshold, due to considerable weight of disconnected clusters.

{\em Effective medium approximation (EMA).}---%
In this approximation one takes into account only the simplest -- local --
correlations between $g_{ij}$ and $R_{ij}$, while all distant
conductances are replaced with a homogeneous medium with effective
conductance $g_{\rm eff}$ (see, e.g., \cite{Kirkpatrick}).
Spatial correlation are neglected within  EMA, and
the system at all ``RG times"  $t$ is completely described by the
single-conductance distribution function $P(g|t)$.
 While being an uncontrolled approximation, EMA provides an
instrument to attack the final stage of evolution of any initial
distribution -- the stage with $\sigma\sim\overline{g}$.
We will see that, as it is typical for EMA, it  works quite well, except
for the immediate vicinity of MIT, for determination of energy-dependent
effective conductance $g_{\rm eff}(t)$.

Within the EMA,
\be
  R_{ij}=\left[g_{ij}+\left(\frac{z}{2}-1\right)g_{\text{eff}}\right]^{-1} .
\label{EMA1}
\ee
The effective
conductance is then found from the self-consistency condition
\cite{Kirkpatrick}
\be
  \langle R_{ij}\left(g_{ij}-g_{\rm eff}\right)\rangle_{g_{ij}} = 0.
\label{EMA2}
\ee
Thus, to find $g_{\rm eff}(t)$ one
should, in principle, solve Eqs.~(\ref{RGE}) and (\ref{EMA1}) with
an arbitrary given $g_{\rm eff}(t)$ and find
$g_{ij}(t)=g[g_{ij}(0),\{g_{\rm eff}\}|t]$ as
a functional of yet unknown function $g_{\rm eff}(t)$, and then,
finally, solve Eq.~(\ref{EMA2}) for $g_{\rm eff}(t)$. This leads to
a nonlinear integral equation
\be
  \int_0^{\infty}
  P_0(g_0) \, dg_0
  \frac{g[g_0,\{g_{\rm eff}\}|t]-g_{\rm eff}(t)}
    {g[g_0,\{g_{\rm eff}\}|t]+\left(\frac{z}{2}-1\right)g_{\rm eff}(t)}=0.
\label{EMA3}
\ee
For a general $P_0(g_0)$ this program can be fulfilled only
numerically. If the distribution $P_0(g_0)$ is narrow, an explicit solution
can be obtained for  $\delta g_{ij}(t) = g_{ij}(t)-\overline{g}(t)$. For the standard deviation
one finds $\sigma_{\rm EMA}(s)/\sigma(0) = e^{-s(1-2/z)}$.
Comparison of this result (for the square lattice case $z=4$)
with the exact perturbation theory (\ref{s/s})
is shown in Fig.~\ref{F:s}. At earlier stages ($s \lesssim 1$) agreement is rather good,
but it becomes worse at large $s$ where ${\bf p}$-dependence of the eigenvalue
$\lambda_-({\bf p})$ becomes important. Thus it seems that EMA may work
reasonably good for broad distributions when $s$ never becomes large.

\begin{figure}
\includegraphics[width=0.9\columnwidth]{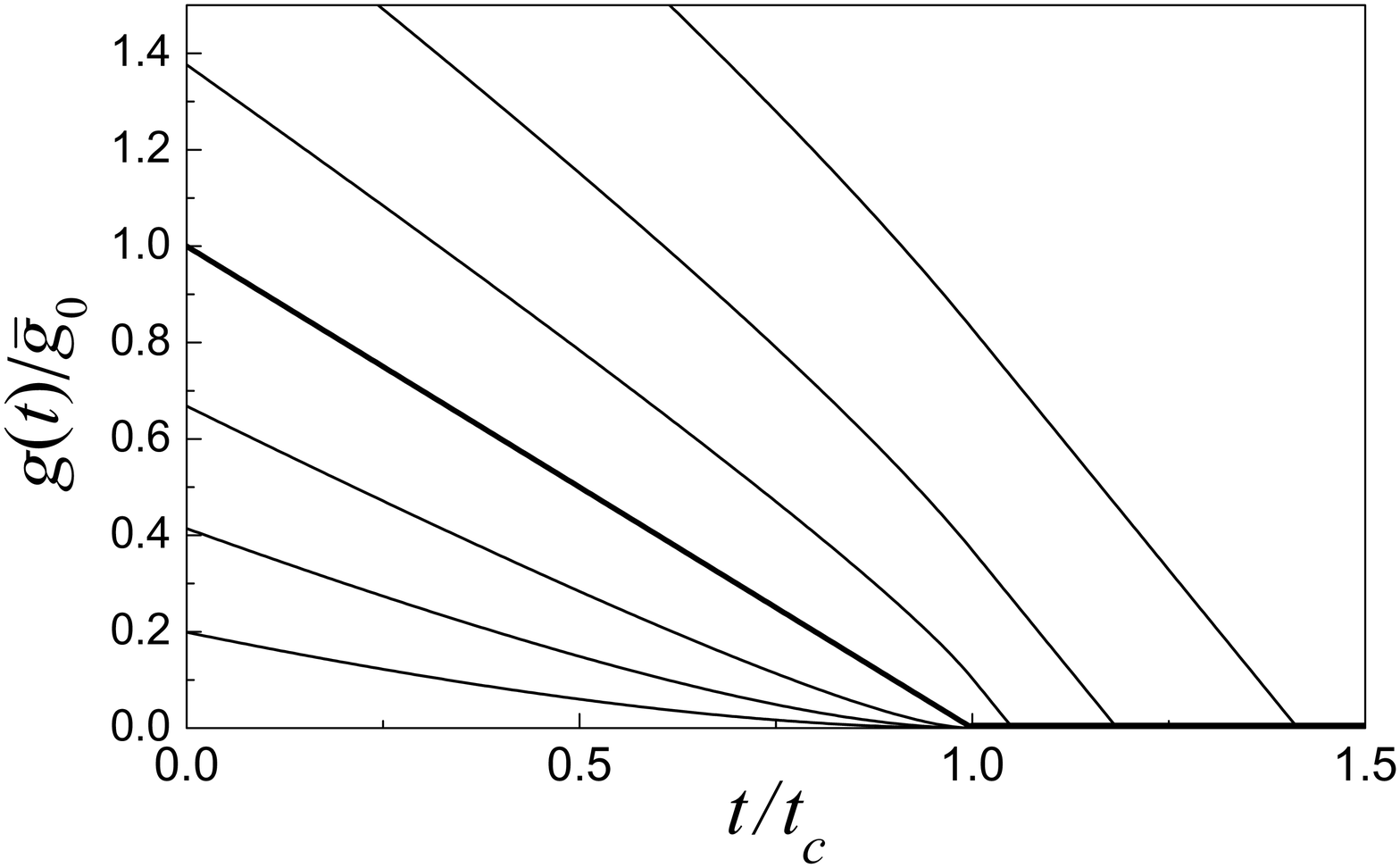}
\caption{Evolution of individual conductances according to the
effective medium approximation. $g_{\rm eff}(t)$ is shown with a thick line.}
\label{F:EMA1}
\end{figure}

An important and physically relevant class of distributions
which allow for analytical EMA treatment is defined by the condition
that $\ln g$ is symmetrically distributed around some mean value.
Writing the tunneling conductance as $g_{ij} = \overline{g}_0e^{-h_{ij}}$
one should require the distribution $p_0(h)$ of fluctuations
$h_{ij} = \kappa (d_{ij} - \bar{d})$, where $d_{ij}$ are
  thicknesses of intergrain insulating barriers, to satisfy $p_0(h) = p_0(-h)$.
For all such distributions on the
square lattice ($z=4$) a simple solution for the effective conductance
can be obtained:   $g_{\text{eff}}(t)=\overline{g}_0-t$
for $t < t_c = \overline{g}_0$
and $g_{\rm eff}(t) = 0$ at $t \geq t_c$.
The individual conductances evolve as follows: for $t<t_c$,
\be
  g_{ij}(t) =
  \left[
    \frac{g_{ij}(0)-\overline{g}_0}{2\sqrt{g_{ij}(0)}}
  + \sqrt{\frac{[g_{ij}(0)+\overline{g}_0]^2}{4g_{ij}(0)}-t}
  \right]^2,
\ee
while for $t>t_c$,
\be
  g_{ij}(t) = [g_{ij}(t_c)-2(t-t_c)] \, \theta [g_{ij}(t_c)-2(t-t_c)].
\label{individi}
\ee
This evolution  is shown in Fig.~\ref{F:EMA1}. The straight line
$g_{\rm eff}(t)$ is the separatrix:  solutions $g(t)$ with
$g(0)>\overline{g}_0$ go above $g_{\rm eff}(t)$ and eventually --
one by one -- become identical zeros at $t>t_c$. Solutions  with
$g(0)<\overline{g}_0$ go below $g_{\rm eff}(t)$ and become zeros
all at once --- at $t=t_c$, together with
$g_{\rm eff}(t)$. For $t_c -t\ll t_c$ the latter solutions form a
narrow ``bunch", manifested by a sharp peak with the total
intensity $\approx 1/2$ in the distribution function $P(g,t)$ at
$g\sim (t_c-t)^2/\overline{g}_0$. For $t>t_c$ this narrow peak
transforms into a $\delta$-peak at $g=0$.
Its intensity $1-N_{\text{cond}}(t)$ [with $N_{\text{cond}}(t)$ being the
fraction of conducting bonds, having $g_{ij}>0$ within the one-loop
accuracy of Eq.~(\ref{RGE})] jumps from $0$ to $1/2$ at $t=t_c$
and then grows monotonically, approaching unity at $t\gg t_c$.

\begin{figure}
\includegraphics[width=0.9\columnwidth]{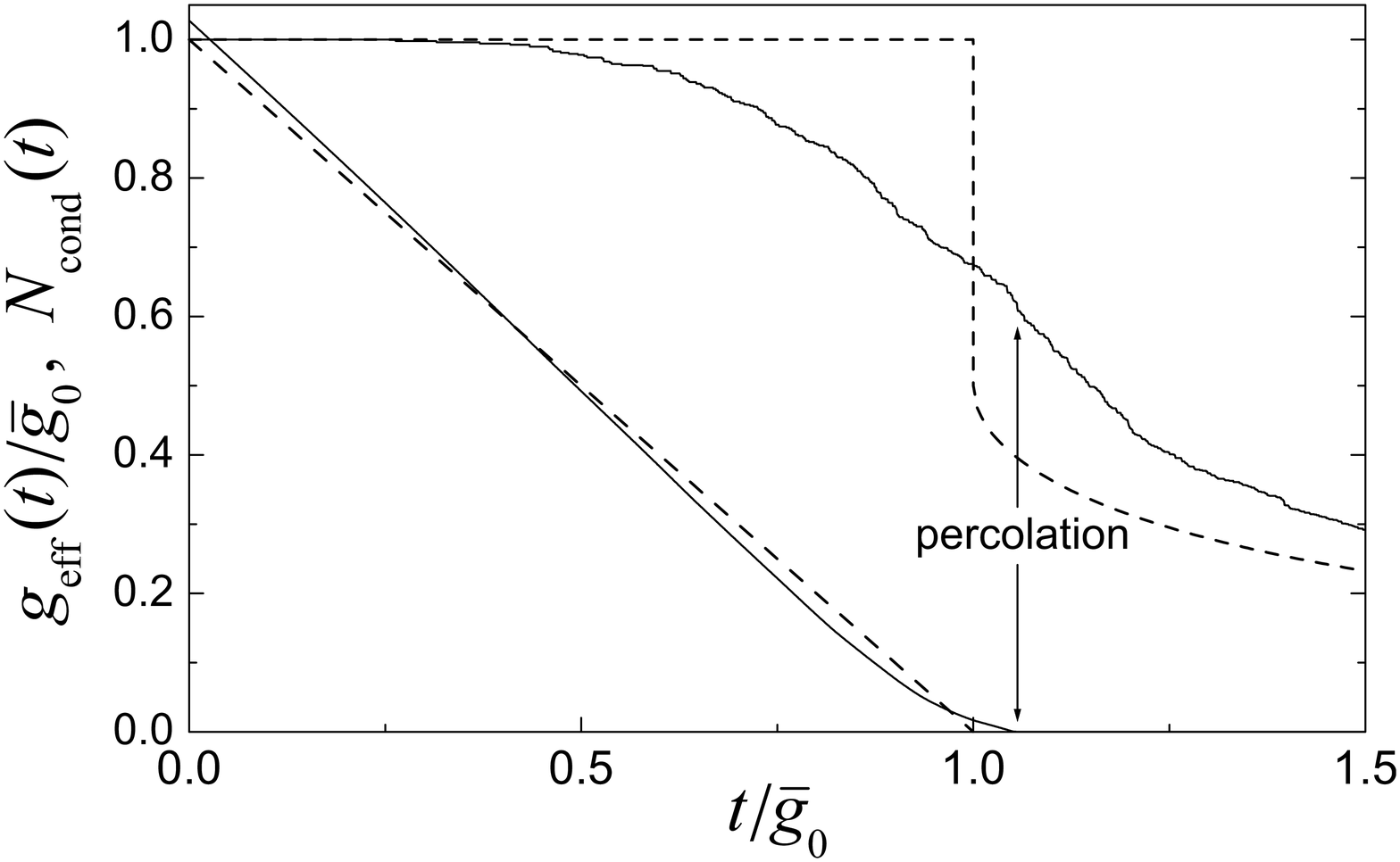}
\caption{Comparison of the EMA (dashed lines) and simulation (solid lines)
results for the global conductivity $g_{\rm eff}(t)$ and the fraction of
conducting bonds $N_{\text{cond}}(t)$ (see text for details).}
\label{F:Lorenz}
\end{figure}

The $t$-dependence of $g_{\rm eff}(t)$ and
$N_{\text{cond}}(t)$ is shown in Fig.~\ref{F:Lorenz} together with the results of
numerical simulations for the Cauchy initial distribution
$P_0(g)=\frac{2\overline{g}_0\theta(g)}{\pi(g^2+\overline{g}_0^2)}$
on the square lattice $32\times32$.
One can see, that the simulated $g_{\rm eff}(t)$ follows the EMA
in the wide range of $t$, while in the vicinity of the transition
it clearly deviates from the EMA and approaches zero with an exponent
$\mu>1$. The numerically found $N_{\text{cond}}(t)$ behaves, however, quite
smoothly, showing no jump at $t=t_c$. The reason for this smoothness is,
apparently the tendency to clusterization, demonstrated
in Fig.~\ref{F:clusters}. It can be interpreted as an instability,
which drives especially small conductances (provided they form
an appropriate configuration) to collapse earlier, than typical ones.

{\em Clustering and percolation.}---%
Let us suppose that, due to a
fluctuation in the initial distribution and/or to the dynamical
evolution up to time $t=t_0$, the conductances $g_{ij}(t_0)$ of all bonds
within a certain group happen to be much smaller, than conductances
in their immediate surrounding. The group should be such that the small
conductances belong to the shell of some cluster of sites, separating
it from the rest of the array (cf.\ Fig.~\ref{F:clusters}).
Then, for any bond $ij$ from the shell one gets
$R_{ij}^{-1}\approx R_{S}^{-1}\equiv\sum_{\rm shell}g_{kl}$.
It means, in particular, that the RG equations for the conductances
of the shell are split from the rest of the RG equations, and can
be solved separately. As a result, one obtains:
\be
  g_{ij}(t) \approx
  g_{ij}(t_0)
  \frac{(t_{\rm shell}-t) \, \theta(t_{\rm shell}-t)}{t_{\rm shell}-t_0},
\label{rg5}
\ee
where $t_{\rm shell}-t_0=R_{S}^{-1}\ll t_c$.
Thus, we conclude, that, in contrast to the EMA solution,
clusters surrounded by poorly conducting shells may become
disconnected from the rest of the array already at $t<t_c$.

Numerical simulation clearly demonstrates the formation of
clusters (see Fig.~\ref{F:clusters}). When a disconnected cluster
appears, the matrix $A$ defined above Eq.~(\ref{dG}) acquires
a new zero eigenvalue. Thus, the total number of electrically decoupled
grains is given by the number of zero eigenvalues of the matrix $A$.
The position $t_c$ of the percolative transition is a functional
of the initial distribution $P_0(g)$. Apparently, $t_c$ is of the order
of some mean initial conductance $\overline{g}_0$, while the correct
coefficient should be determined numerically.

In general, evolution of the network of conductances with the growth
of the parameter $t(E)$ is rather similar to that would be expected
at the classical percolation transition. However, our system can not
be described by either purely ``bond" or ``site" percolation,
due to development of local correlations (clustering) along with the RG flow.
In particular, numerically observed (cf.\ Fig.~\ref{F:Lorenz})
value of $N_{\text{cond}}(t_c)$ is clearly larger than $1/2$,
contrary to expectations for purely bond percolation in 2D.
More detailed numerical work is needed to determine the nature of this
new kind of percolative transition; in particular, ``measurements" of the
conductivity exponent $\mu$ (equal to 1.3 in the standard percolation
problem~\cite{Grassberger}) would be very desirable.

{\em Conclusions.}---%
We have shown that at low temperatures strong intrinsic
inhomogeneities are developing in granular metal arrays with
moderately large random  bare conductances $g_{ij} \gg 1$.
As a result, the Coulomb-driven metal-insulator transition
expected if $g_0 \leq \ln(E_C/\delta)$~\cite{bel1,kam04}
acquires features of percolation transition. Most directly the predicted
behavior can be detected by measuring the distribution of the local
tunneling density of states at low temperatures.
The best object for such a study would be a granular cermet
of metal grains in the insulating matrix, like those studied
in Refs.~\cite{gerber,simon}. In these materials
the ratio $E_C/\delta$ was about $10^3$, indicating
the existence of a broad range for logarithmic corrections to conductivity.
It is hardly possible that local tunnel conductances in such a
granular cermet are all equal; at best, they can be distributed
with the width of the order of the mean conductance. Our results
presented in Fig.~\ref{F:Lorenz} show that a simple logarithmic
dependence $g_{\rm eff}(T) = \overline{g}_0 - \ln(\overline{g}_0E_C/T)$
holds in a wide range of $T$ for moderately random granular arrays as well,
at least for the class of practically important symmetric distributions
of $\log(g)$ in the 2D space.

If a granular metal has a tendency to become superconductive
with $T_{\rm sc} \sim T_c$, its local superconductive properties are
expected to be strongly inhomogeneous due to position-dependent
Coulomb effects. In other terms, superconductive properties of
granular metal can be much more of ``granular nature" than its
normal properties at elevated temperatures. In this regard
we mention very interesting recent experimental results~\cite{Walter}.

We are grateful to I. S. Beloborodov, A. M. Finkelstein, A. V. Lopatin,
and V. M. Vinokur for useful discussions on initial stage of this project.
This research was supported by the Program ``Quantum Macrophysics" of
the Russian Academy of Sciences, Russian Ministry of Science, RFBR under
grants No.\ 04-02-16348 and 04-02-16998, and by the US Department of Energy,
Office of Science through contract No.\ W-31-109-ENG-38.
M.~A.~S. acknowledges financial support from the Dynasty foundation
and the ICFPM.


\begin{thebibliography}{99}
\bibitem{comment-g}
Note that in Refs.~\cite{efetov,bel1,bel2} the dimensional conductance
$g$ is normalized to $2e^2/\hbar$, i.e. it differs by the factor
$4\pi$ from the standard definition we use.
\bibitem{comment-A}
The matrix $A$ has the zero eigenvalue associated with the shift
of all voltages by a constant. Thus, the inverse matrix $G$ does not exist.
However, expression for $R_{ij}$ remains well defined since it does not
contain the zero mode.
\end{thebibliography}
\end{document}